\documentclass[11pt, a4paper]{article}
\usepackage{jheppub}

\usepackage{amsmath,amssymb}

\newcommand{\E}{\mathbb{E}}

\DeclareMathOperator{\sign}{sgn}
\makeatletter
\def\@fpheader{\relax}
\makeatother

\title{Bohmian Mechanics fails to compute multi-time correlations}
\author{Robert C. Helling\\
Arnold Sommerfeld Center\\
Ludwig-Maximilians-Universit\"at M\"unchen\\
Theresienstra\ss e 39\\
80333 M\"unchen\\
Germany\\
{\tt E-mail: helling@atdotde.de}}

\abstract{The violation of Bell type inequalities in quantum
  systems manifests that quantum states cannot be described by
  classical probability distributions. Yet, Bohmian mechanics is a
  realistic, non-local theory of classical particle trajectories that is claimed
  to be indistinguishable by observations from more traditional
  approaches to quantum mechanics. We set up a spatial version of the GHZ
  system with qubits realised as positional observables that
  demonstrates that the Bohmian theory fails to match predictions of
  textbook quantum mechanics (and most likely experients)
  unless enlarged by a microscopic theory
  of collapse of the wave function after observation. For
  this discrepancy to occur it is essential that positions at
  different times do not commute.}

\begin{document}
\maketitle

\section{Introduction}
\setcounter{equation}{0}

The announcement of the 2022 Nobel Prize in Physics to  Alain
Aspect, John Clauser and Anton Zeilinger for the experimental
demonstration of the violation of Bell type inequalities has sparked
another round of arguments of what this actually implies for quantum
theory.

In particular supporters of the Bohmian approach\cite{Duerr} instited that by no
means hidden variable theories are ruled out as their example of a
hidden variable theory (with the Bohmian particle positions playing
the role of the hidden variables) reproduces all the experimental
predictions of more traditional approaches to quantum mechanics. It
does this at the price of being non-local. Some have even argued that
the Bell violation implies that quantum mechanics has to be a
non-local theory.

In this note, we will examine this claim that the Bohmian theory
matches all predictions of traditional quantum mechanics in more
detail. To this end, we will study a version of the GHZ experiment (in
the version popularized by Mermin and Coleman \cite{mermin,coleman}) which is close in
spiri to the Bell inequality but has the advantage of making definite
statements instead of giving probabilities for measurement outcomes. We
will, however, not formulate it in terms of spin degrees of freedom
but rather use simple positional observables.
Those are much better suited to a Bohmian description as the Bohmian
theory in its original form
posits all measurements in the end are measurements of positions:
These can be positions of a pointer on a scale or in the case of spin
degrees, positions of particles for which a spin degree has been
translated in terms of a Stern-Gerlach-type experiment to a
trajectory. By directly working with positional observables, we evade
the complication of having to include the Stern-Gerlach set-up in our
description including a Hamiltonian and an explicit solution for a
time dependent wave function.

In order to realize the violation of the quantum inequalities, one has
to make use of entanglement which in turn requires non-commuting
observables. This seems to be in conflict with considering only
position observables. But this can be circumvented by measuring
positions at different instances of time as postions at different times
are no longer required to commute. It is important though that the
measurements at the different times are with respect to different
particles or degrees of freedom and thus we do not attempt to measure
complementary variables. It is only that for each degree of freedom we
have the freedom to choose the time of measurement and different times
correspond to observables that do not commute. But we always only
choose one of those possibilities.

What we find is that the Bohmian approach fails to reproduce the
expected results unless earlier observations influence later ones by the
equivalent of a collapse of the wave function. After the first
observation of one particle, the other Bohmian particles have to move with respect to a
collapsed wave function rather than the original one. Thus, in order to be
predictive, there has to be a quantitative understanding of this
collapse process which is not in sight.  The claim
of equivalence of the two approaches can only hold in a Bohmian theory
that is augmented by a mechanism that leads to a collapsing wave
function. To be able to make quantitative predictions (that can be
compared to the concrete predictions of more conventional approaches to quantum
mechanics) it is not enough to simply state that some sort of collapse
happens (or that somehow the wave function of the measuring device
influences the system) but concrete equations of motion have to be
supplied. If the collapse is not to be simply postulated in an ad hoc way, a
microscopic mechanism for such a collapse is needed.

\section{Setting the stage}
Key to the Bohmian description of quantum mechanics is the
conservation of the probability current for Hamiltonians of the
form
\
\begin{equation*}
  H = -\sum_{i=1}^N \frac 1{2  m_i}\frac{\partial^2}{\partial x_i^2}+ V(x_1,\ldots, x_N) 
\end{equation*}
for a system with $N$ degrees of freedom where $m_i$ is the mass of
the $i^{\rm th}$ degree of freedom.
Dividing the probability current by the
probability density leads to a velocity field
\begin{equation*}
  v_i(t; x_1,\ldots , x_N) =\frac{j_i(t, x_1,\dots, x_N)}{ |\psi(t,
    x_1,\ldots, x_N)|^2}= \frac 1{m_i}\Im \left(\frac{\partial\psi(t; x_1,\ldots,
    x_N)}{\partial x_i}\Big / \psi(t, x_1,\ldots, x_N)\right).
\end{equation*}
The de Broglie-Bohm theory posits now microscopic particles with
positions $q_i(t)$ that follow the flow of this velocity field:
\begin{equation}
  \label{eq:guidingeq}
  \frac{dq_i}{dt} = v_i(t; q_1(t),\ldots,q_N(t)). 
\end{equation}
If at time zero, these particles are distributed according the
probability density
$$|\psi(0, x_1,\ldots, x_N)|^2$$
they will, thanks to
the conservation of the probability current, be so at any other time
as well. Thus, if all one is interested in are the distribution of the
positions of particles (which is argued is all one would ever
measure), the particle distribution of these microscopic particles is
as predicted by traditional quantum mechanics.

It is important to realize though that the guiding equation
(\ref{eq:guidingeq}) is non-local in the sense that the velocity of
degree of freedom $i$ depends on the values of all other degrees of
freedom, even if these are the coordinates of different particles that
can be at arbitrarily large spatial separation. This non-locality is usually
invoked in order to avoid the conclusions of Bell's theorem as that
assumes a local theory in the first place. Without locality, the
outcome of a measurement on one particle can, at least in principle,
depend on the settings of all detectors in the experiment. Our
analysis will show, however, that this non-locality is not sufficient
to produce the correlations in the GHZ experiment.

The positions $q_i(t)$ are the ``realistic'' state in the sense that
the particles are objectively ``there'' even when not observed like
for example the true microstate of a (classical) gas when only a
macroscopic state is prescribed.

On the other hand, Bell's inequality (and the likes like the GHZ experiment
we are concerned with here) are derived assuming a local, realistic
theory. Their violation in the quantum theory thus shows that the
quantum world cannot be both local and realistic.

The GHZ system is tri-partite. On each of the three sub-systems (which
can be thought of as separated by an arbitrarily large spatial distance) there are
two possible measurements both with outcomes either $+1$ or $-1$. We
will denote these two measurements of sub-system  $a$ (and in slight
abuse of notation also the outcome and later the corresponding
observable) $X_a$ and $Y_a$ and similarly for sub-systems $b$ and $c$.

The whole GHZ experiment is repeated over many rounds. In the
beginning of each round, the tri-partite system is prepared in the
same particular ``GHZ-state''. Then the three experimentators at the
sub-systems randomly pick which of the two possible measurements they
will conduct. They do the observation and write down which measurement
they performed and the $\pm 1$ outcome.

Later, they compare their results. They find the outcomes of individual
measurements to be random with vanishing expectation. Also all
correlations of two of the measurements at two sub-systems are found to
vanish. But there are non-trivial correlations when all three
measurements are taken into account: They find that if they all did
the $X$ measurement the product of the outcomes was always $-1$:
\begin{equation}
  \label{eq:XXX}
  X_a X_b X_c = -1
\end{equation}
In the case of one experimentator measuring $X$ and the other two operators
measuring $Y$ the product is always $+1$:
\begin{equation}
  \label{eq:XYY}
  X_aY_bY_c = Y_aX_bY_c = Y_aY_BX_c = +1
\end{equation}
Since all outcomes are $\pm 1$, these equations can also be written as
\begin{equation}
  \label{eq:Yeq}
  X_a = Y_bY_c\qquad  X_b=Y_aY_c \qquad X_c=Y_aY_b.
\end{equation}
In other words, if the measurements measure actual, preexisting
properties of the sub-systems (i.e. that we are dealing with a realist
theory), we can conclude that in the original state the $X$-property
of one sub-system is determined by and can be deduced by the
$Y$-properties of the other two sub-systems. 

But taking the product of these equations one finds
\begin{equation}
  \label{Xprod}
    X_aX_bX_c = (Y_aY_bY_c)^2 = +1,
  \end{equation}
  In clear conflict with (\ref{eq:XXX}): Let's assume both $X_a$ and
  $X_b$ both measure $+1$. Then (\ref{eq:XXX}) implies that $X_c$ has
  to be $-1$ while (\ref{Xprod}) impies that $X_c$ at the same time has to be $+1$.

  This contradictions seems to
  make any such experiment with this type of correlations of outcomes
  impossible, at least in a local theory. The physical property that
  $X$ measures (which we will identify with particle positions below)
  cannot be both anti-correlated (as suggested by (\ref{eq:XXX})) and
  correlated (as suggested by taking the three permutations of
  (\ref{eq:XYY} into account) at the same time. In a non-local theory, however, the
  three sub-systems can ``know'' which of the measurements are
  conducted on the other sub-systems and adjust their outcome
  accordingly. 

 Quantum mechanically, by making use of entanglement it is possible to evade the
 impossibility: For example, one can assume the subsystems contain a
 qubit each and the $X$ and $Y$ measurements are the corresponding
 Pauli spin operators:
 \begin{equation}
   \label{Pauli}
   X=
   \begin{pmatrix}
     0&1\\
     1&0\\
   \end{pmatrix}
   \qquad
    Y=
   \begin{pmatrix}
     0&i\\
     -i&0\\
   \end{pmatrix}
 \end{equation}
 For the initial state, GHZ take
 \begin{equation}
   \label{GHZstate}
   \Psi_{GHZ} = \frac 1{\sqrt 2}\big(|\uparrow \uparrow 
\uparrow\rangle- |\downarrow\downarrow\downarrow\rangle\big)
 \end{equation}
 Applying $X$ to all three qubits flips them and $\Psi_{GHZ}$ is
 clearly an eigenstate of eigenvalue $-1$. Applying one $X$ and two
 $Y$-operators adds two factors $(\pm i)^2=-1$ in addition to the flip
 and thus $\Psi_{GHZ}$
 is a $+1$ eigenstate of $X_aY_bY_c$,  $Y_aX_bY_c$,  and $Y_aY_BX_c$.

 One (in this authors opinion: the) way to reconcile quantum theory with the above derivation of
 (\ref{Xprod}) is to give up realism: Since $X$ and $Y$ do not
 commute, they are complimentary variables that cannot both be
 measured at the same time on the same sub-system. Thus in a
 calculation involving $X_a$, it is not valid to assume that $Y_a$ has
 any (unknown but somehow objectively existing) value and similar for
 the other two sub-systems. So it is not legitimate to write an
 expression for $X_aX_bX_c$ that involves the $Y$'s since $Y_a$
 fundamentally cannot be measured as well besides $X_a$ and thus one must not assume
 they still have a value. This is the local but non-realist way to
 explain the observed quantum correlations that appear to be
 impossible in a classical (and thus realist) theory.

The Bohmian theory goes in a different direction: There one insists on
realistic degrees of freedom (the $q_i$'s) but is willing to give up
locality. In this note, we will investigate how this is supposed work in
detail and if the conflict between (\ref{eq:XXX}) and (\ref{Xprod}) is
resolved in the Bohmian theory.

\section{The Bohmian GHZ}
Even though there is a version of the Bohm theory that can handle spin
degrees of freedom, following the mantra of ``all measurements in the
end measure positions'' one is supposed to run those through a
Stern-Gerlach apparatus or similar to translate spin states into postion
states. This complicates the analysis significantly since it means one
has to take into account the Stern-Gerlach apparatuses in the systems
Hamiltonian and guiding equation.

It is even argued\cite{ops1,ops2} that the only true properties of
particles are their positions while spin is rather a property of the
wave function and the observed value depends on the details of the
experiment that is supposed to measure it. This explicitly allows for
contextualism in the way that the result of the spin measurement on
one particle (in terms of a Stern-Gerlach apparatus) depends on which
spin component of a different particle is measured (in terms of
setting up another Stern-Gerlach apparatus). After all, according to
\cite{ops1,ops2}, only positions $q_i$ re both real and measurable.

But after all, all that the above description really needs are
binary outcome observables $X$ and $Y$ that fulfil the required
commutation relations. Two state systems are abundant also in spatial
systems. But non-commuting observables might seem impossible to obtain if all one
is allowed to use are multiplication operators in position
representation. One can even speculate that restriction to such
commuting position operators is what makes it possible for a theory
that contains realistic particle orbits to behave like quantum theory.

Still, it is possible to avoid this restriction taking into account
time evolution as positions at different times no longer
commute. Specifically, we are going to use a set-up similar to what
was used in \cite{bohmianangels} which was inspired by
\cite{werner,CorreggiMorchio}.  A related situation with two time
correlations was also investigated by Gisin in \cite{Gisin} as well as
\cite{laloe, neumaier}.

\begin{figure}
  \centering
\includegraphics{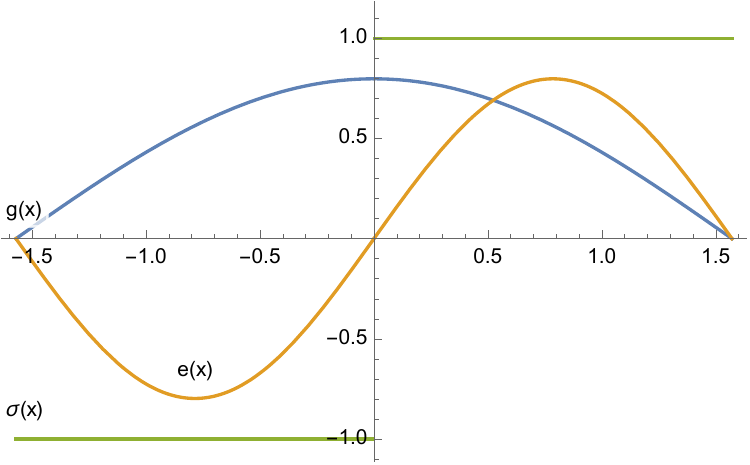}  
  \caption{Wave functions of ground state and first excited state as
    well as $\sigma$}
  \label{fig:wavefcts}
\end{figure}

Let each sub-system consist of a particle in a box which we take to be
the interval $[-\pi/2, \pi/2]$ and thus the Hilbert space as
$L^2([-\pi/2,\pi/2])$ with the Hamiltonian of the free particle $H=
-\frac{d^2}{dx^2}$ where we have set all masses $m_i=1/2$ for
simplicity. We impose Dirichlet boundary conditions. We focus our attention to the ground state and the
first excited state
\begin{equation}
  \label{wavefunctions}
  g(x) = \sqrt{2/\pi}\cos(x)\qquad e(x) = \sqrt{2/\pi}\sin(2x)
\end{equation}
with respective energies $E_g=1$ and $E_e=4$. In this system, we ask if
the particle is in the left or in the right half of the interval which
corresponds to the observable given by the multiplication operator by
$\sigma(x) = \hbox{sgn}(x)$ which is $-1$ in the left half and $+1$ in
the right half, see Fig.~\ref{fig:wavefcts}.

Since $g(x)$ is an even function while $e(x)$ and $\sigma(x)$ are odd,
the expectation of $\sigma$ in both states vanishes
\begin{equation}
  \label{sigmaex}
  \langle g|\sigma|g\rangle = \langle e|\sigma|e\rangle=0, 
\end{equation}
but there is a non-trivial overlap
\begin{equation}
  \label{overlap}
  \langle g|\sigma|e\rangle = \frac 8{3\pi}\approx 0.85
\end{equation}
This was all at the initial time $t=0$. As they are energy
eigenstates, $|g\rangle$ and $|e\rangle$ simply time evolve with a
phase, so we have
\begin{equation}
  \label{overlapattime}
  \langle g|e^{-iHt}\sigma e^{iHt}|e\rangle = \frac 8{3\pi}e^{i\Delta Et}
\end{equation}
In the sector spanned by these two states, the $\sigma$-operator in
the Heisenberg-picture $\sigma_t=e^{-iHt}\sigma e^{iHt}$ becomes
\begin{equation}
  \label{heisenberg}
  \sigma_t=\frac 8{3\pi}
  \begin{pmatrix}
    0&e^{i\Delta Et}\\
    e^{-i\Delta Et}&0\\
  \end{pmatrix}.
\end{equation}
So, up to a numerical factor, $\sigma_0$ plays the role of $X$, the Pauli
spin matrix in the $x$-direction while $\sigma_{\pi/(2\Delta E)}$ is
$Y$, the
spin matrix in the $y$-direction. We have found a convenient set of non-commuting observables.

We can now set up the analogue of the GHZ-state: We use three copies
of the the particle in a box $L^2([-\pi/2,\pi/2])\otimes
L^2([-\pi/2,\pi/2])\otimes L^2([-\pi/2,\pi/2]) \cong
L^2([-\pi/2,\pi/2]^3)$. This system can equivalently be viewed as
three particles confined to three separate intervals (which can be
thought of as located at large spatial separation) or alternatively as one
particle in a three-dimensional box.

\begin{figure}
  \centering
  \includegraphics[width=15cm]{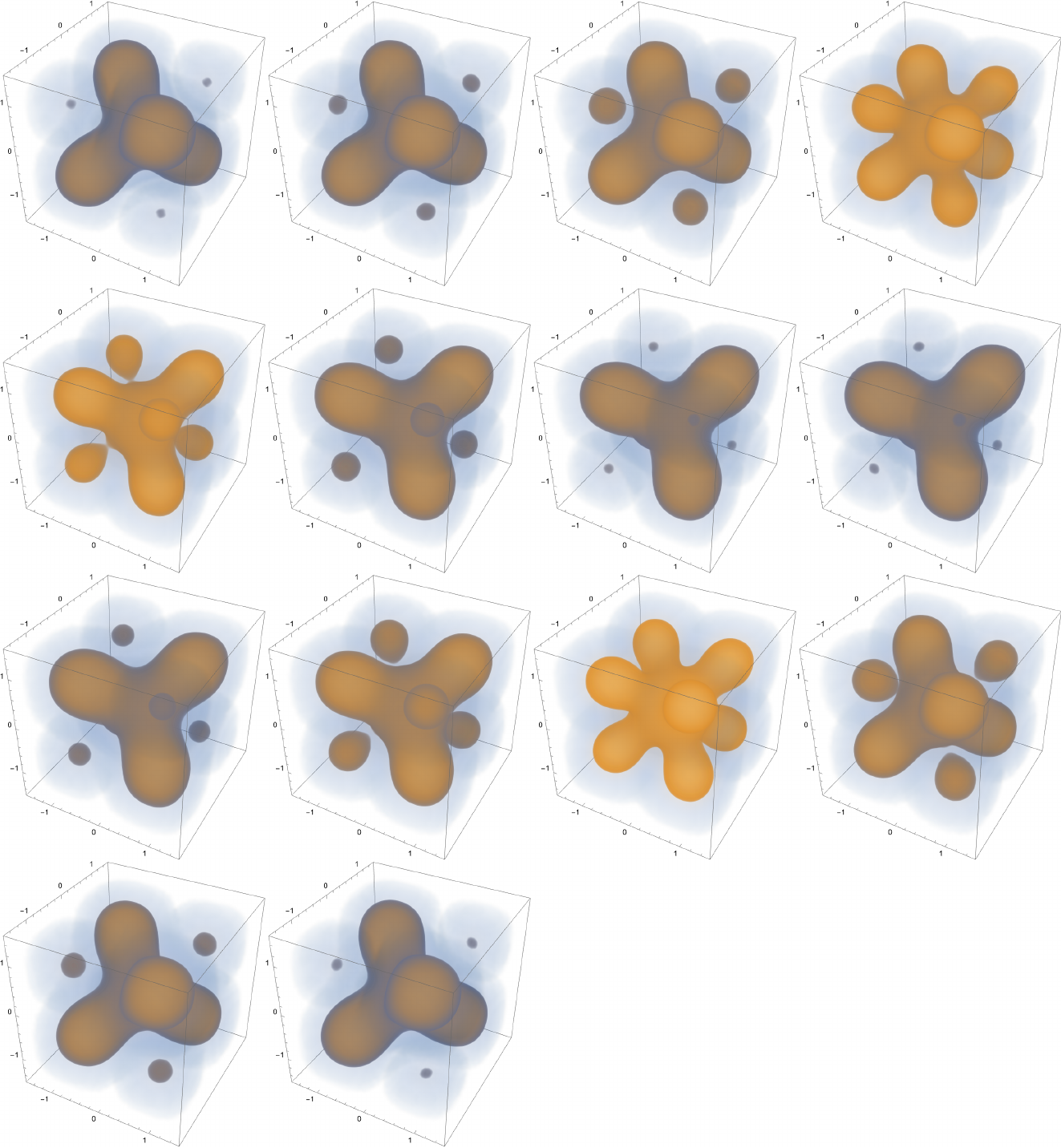}
  \caption{A comic of the periodic time evolution of the probability density
    $|\Psi_{GHZ}|^2$ as a particle in a three-dimensional box}
  \label{fig:comic}
\end{figure}

We set
\begin{equation}
  \label{ghzstate}
  \Psi_{GHZ}= \frac 1{\sqrt 2} \left( g\otimes g\otimes g - e\otimes
    e\otimes e\right)  
\end{equation}
or
\begin{equation}
  \label{ghzstatecoord}
  \Psi_{GHZ}(x,y,z)= \frac 1{\sqrt 2} \left( g(x) g(y) g(z) - e(x)
    e(y)e(z)\right)    
\end{equation}
Different from the state considered in \cite{bohmianangels}, $\Psi_{GHZ}$ is not stationary with respect to the free particle time
evolution. The time evolution of the probability density is shown in Fig.~\ref{fig:comic}.

\begin{figure}
  \centering
  \includegraphics[width=7cm]{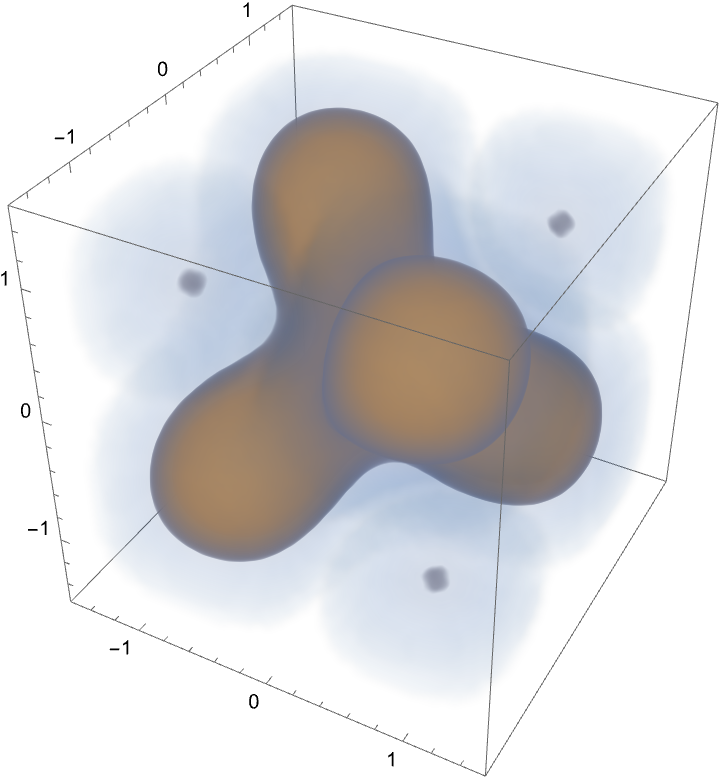}\includegraphics[width=7cm]{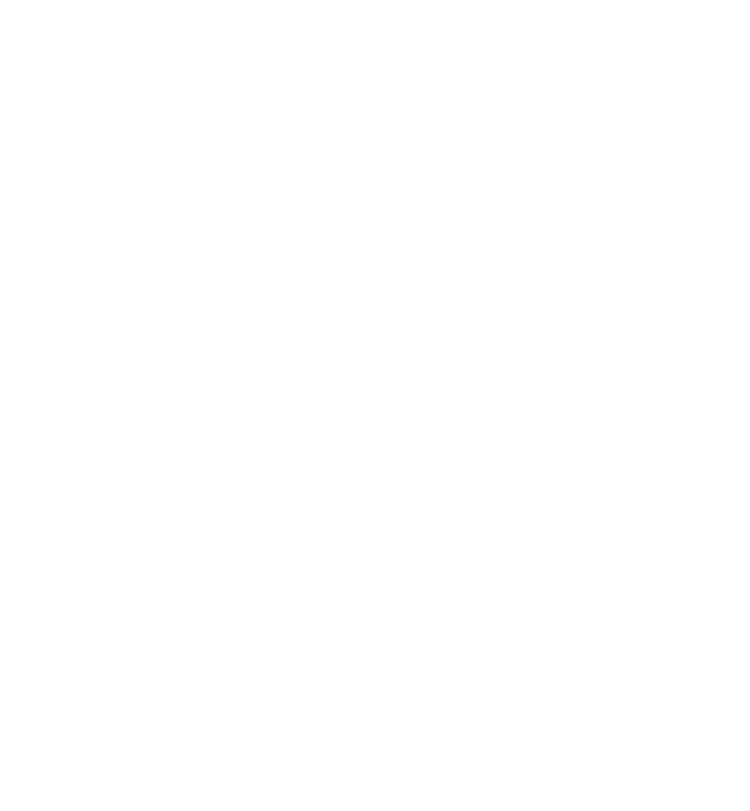}
  \caption{Propability distribution at $t=0$ and $t=\pi/(2\Delta E)$}
  \label{fig:prob0}
\end{figure}
At $t=0$ (Fig.~\ref{fig:prob0}), we can ask if the three particles are in
the left or the right half of the interval. What is apparent from the
probability distribution Fig.~\ref{fig:prob0} can also be computed:
\begin{equation}
  \label{eq:tnullcorrelation}
  \langle
  \Psi_{GHZ}|\sigma\otimes\sigma\otimes\sigma|\Psi_{GHZ}\rangle= -\left(\frac 8{3\pi}\right)^3
\end{equation}
Up to a numerical factor, this is nothing but (\ref{eq:XXX}) in this
positional setting. In other words, the initial probability
distribution of the Bohmian positions as implied by $\Psi_{GHZ}$ has a
strong anti-correlation: In the language of a particle in a
three-dimensional box, the majority of the probability is in four out
of the eight possible octants.

Similarly, we can do this at times $s$, $t$ and $u$ for the respective
particles. A short calculation shows
\begin{equation}
  \label{eq:stucorrelation}
  \langle
  \Psi_{GHZ}|\sigma_s\otimes\sigma_t\otimes\sigma_u|\Psi_{GHZ}\rangle=
  -\cos(\Delta E(s+t+u))\left(\frac 8{3\pi}\right)^3.
\end{equation}
In particular, to find the expression corresponding to (\ref{eq:XYY}), we
evaluate this for two times taken to be at $\pi/(2\Delta E)$ and the last one at
$0$ (our set-up is of course invariant under permutations of the three
particles):
\begin{equation}
  \label{eq:xyycorrelation}
  \langle
  \Psi_{GHZ}|\sigma_{\pi/(2\Delta E)}\otimes\sigma_{\pi/(2\Delta E)}\otimes\sigma_0|\Psi_{GHZ}\rangle= +\left(\frac 8{3\pi}\right)^3.
\end{equation}

So picking any two of the three particles, compared to the initial
position at $t=0$, exactly one of the pair has to have changed the side in
the box at time $\pi/2\Delta E$. Of course, this is not possible for
all possible pairs of two out of three particles!

So, up to a fidelity factor of $(8/3\pi)^3$, we recovered the
situation of the GHZ experiment: From the three possible correlations
of the positions of one particle at time zero with the two other
particles at time $\pi/(2\Delta E)$, we would conclude by the same argument that
lead to (\ref{Xprod}) that more likely the
product of the signs of the positions of the three particles at the
initial time is positive while it is actually negative.

Or put differently: The correlation (\ref{eq:stucorrelation}) is not
reproduced by the positions $q_i$ of the Bohmian particles if they
evolve according the the evolution (\ref{eq:guidingeq}). On the other
hand, anything else would have been surprising as the position of a
particle would have to be both on the left as well as the right half
of the interval depending on if one argues in analogy to (\ref{eq:XXX})
or (\ref{Xprod}). That is of course impossible for ``objective''
particle positions.

\begin{figure}
  \centering
  \includegraphics[width=7cm]{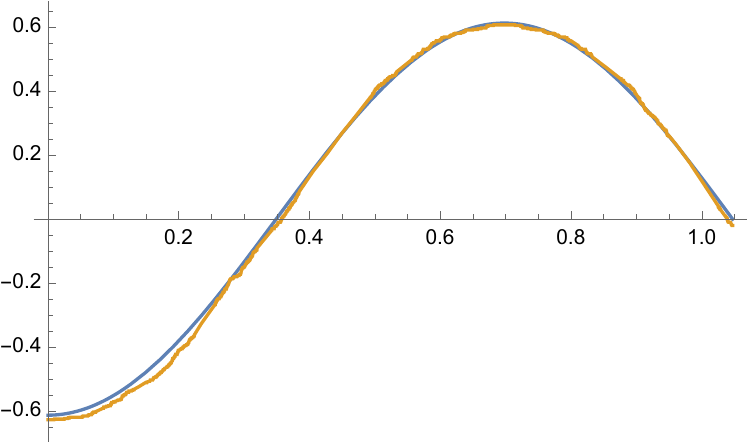}
  \caption{(\ref{eq:stucorrelation} (blue) for $t=s=u$ and the Bohmian expression
    $\E\langle\sign(q_1(t))\sign(q_2(t))\sign(q_3(t))\rangle$ (orange)
    agree well}
  \label{fig:equaltimes}
\end{figure}

To strengthen this claim, we picked 1000 sets of initial positions from the
probability distribution $|\Psi_{GHZ}|^2$ and numerically integrated
the guiding equation
(\ref{eq:guidingeq}) using mathematica. In the special case $s=t=u$ where the positions of
the three particles are evaluated at identical times and the
corresponding observables thus commute, we find excellent agreement
with the expression (\ref{eq:stucorrelation}) as shown in
Fig.~\ref{fig:equaltimes}.

\begin{figure}
  \centering
  \includegraphics[width=7cm]{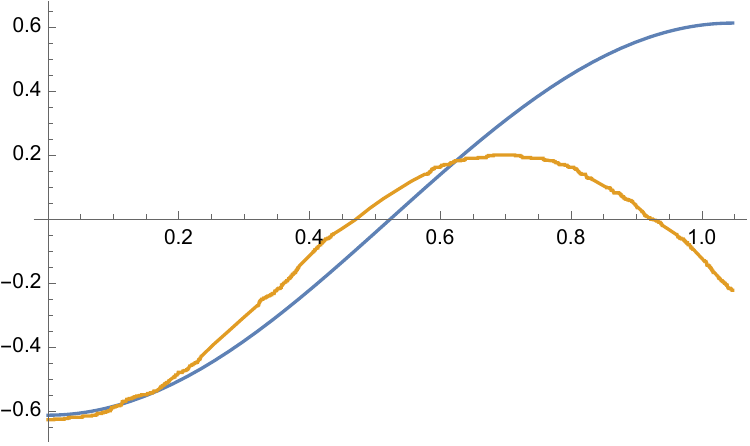}
  \caption{(\ref{eq:stucorrelation}) for $t=0$ and $s=u$ (blue) and the
    Bohmian expression for the particles
    $\E\langle\sign(q_1(0))\sign(q_2(s))\sign(q_3(s))\rangle$ (orange)
    no longer agree}
  \label{fig:twotimes}
\end{figure}

If, however, we evaluate the position of one particle at the initial
time $t=0$ the other two particles' position at $s=u$, the Bohmian
expression no longer agrees with (\ref{eq:stucorrelation}) as is clear from
Fig.~\ref{fig:twotimes}. We find that correlations of the $q_i$ at
different times do no longer agree with predictions for correlations
of the corresponding Heisenberg picture observables. Positions at
different times are not covered by the general {\sl ``the Bohmian particle
positions are distributed according to $|\psi|^2$ at every time and
thus the pilot wave theory automatically reproduces all predictions of
traditional quantum mechanics''} explanation.

Even the non-local nature of the guiding equation is not able to
reconcile the particle positions with the quantum theory predictions.

\section{Ways out}
There is, however, a way to save the Bohmian theory from this
discrepancy and give it a chance to reproduce the conventionally
computed result (which we strongly believe would be confirmed
experimentally). We leave it to the reader to judge by how much this
weakens the explanatory power of the Bohmian predictions.

This route starts with the realisation that it is not sufficient to
ask where the three particles ``are'' at the given points in time
(which would be expressed by the $q_i$ as they are supposed to be
thought as the {\sl actual} particle positions that would for example
blacken exactly one silver grain on a photographic plate) but that it
would have to be actually measured. But each measurement is known to
disturb a quantum system.

We should point out that in the literature\cite{ops1,ops2} on Bohmian
mechanics it is stressed that the Bohmian theory of trajectories is
concerned with ``what is really the case'' not just with what some
complicated apparatus is supposed to show. In fact, the focus of other
approaches on observables is frowned upon since observables and
measurements require an observer and thus introduce a subjective
notion (what actually constitutes an observer?) into the
discussion. The Bohmian theory in contrast claims to be about quantum
theory without observers.

But leaving this attitude that has potential to challenge the nature
of physics being an empirical theory (where certain notions like the
particle positions $q_i$ are claimed to exist while at the same time
are denied to be empirically accessible), one can entertain the idea
that ``where are the three particles at different times?'' is not the
right question but rather ``where can I measure the particles at
different times?''. In \cite{Gisin}, after a similar analysis, Gisin
proposes to give up what he calls ``Hypothesis H'' which states the
strong connection between the Bohmian particle positions and the
outcome of position measurements. This conclusion, however, severs the
the Bohmian position variables completely of empirical content.

So if one decides to measure the position
$q_1$ of the first particle at $t=0$, from that point on, one should no
longer expect that the other two particles follow the guiding equation
derived from $\Psi_{GHZ}(t)$ but rather from a collapsed wave
function. Or said differently, because of its non-local nature, the
guiding equation is not supposed to be applied to sub-systems, only to
the wave function of the entire universe which not only contains the
three particles with their postions but at least the measurement
device used to detect the position $q_1$ of the first particle at
$t=0$. Taking this seriously, however, means  one can never be sure
what the actual guiding equation is unless one has control over all
degrees of freedom in the universe. There are no decoupled or approximate
sub-systems and the predictability of the theory in any realistic
situation is in doubt. Anybody who wants to argue along these lines
should provide a calculation within the Bohmian framework which is
able to reproduce the blue curve in Fig.~\ref{fig:twotimes} which is
an experimentally testable prediction of the orthodox theory. The
failure to compute this curve or the the claim that the outcome
depends on further so far unspecified details of the experimental
set-up constitute a lack of predictable power of the theory.

But this seems to be the view at least of some supporters of the
Bohmian theory, at least that is our understanding of discussions in
social media.\footnote{Specifically comments solicited on this question
  posed on https://atdotde.blogspot.com/2024/05/what-happens-to-particles-after-they.html}

Another problem for a Bohmian model of a measuring device is that only
the guiding equation (\ref{eq:guidingeq}) is sourced by the wave
function but on the other hand there is no feedback from the particle
positions $q_i(t)$ on the wave function which evolves according to the
Schr\"odinger equation independently of the $q_i(t)$. So no detector wave function can depend on the
$q_i$ of the measured particles. So within the Bohmian theory, it is
not possible to describe dynamics of a measuring device which picks up
the ``actual positions'' of the particles that it is supposed to
measure. The observation of the values $q_i(t)$ can only happen
outside of the Bohmian description.

It is important to stress that for other physical theories it is not
the case, that since ``everything is connected with everything'', it
is impossible to make predictions for the correlations of the
positions of the three particles at different times without detailed
knowledge of the detection mechanism. In fact, standard quantum
mechanics makes these predictions (\ref{eq:stucorrelation}) for this
system without additional knowledge of the microscopic details of the
measuring device. So to match those predictions, the physics (as far
as relevant for these predicted outcomes) must not depend on those
details of the measuring device: Only one outcome can be correct, that
one that matches the predictions of standard quantum mechanics.

Alternatively, one could try to implement some sort of collapse in the
moment where the position of the first particle is detected. One has
to throw away those possible paths $q_i(t)$ that are not compatible
with the measurement and from the time of the measurement use a
collapsed wave function
\begin{equation*}
  \frac{P\Psi_{GHZ}}{\|P\Psi_{GHZ}\|}
\end{equation*}
in the guiding equation where $P$ is the projector on the eigen-space
of the measured observable (with unclear meaning in the case of
continuous spectrum).

In any case, more detailed information (the form of the correct
projector $P$ above) is required in order to obtain
an updated  guiding equation after the collapse in order for the
theory to be able to reproduce the relatively easy prediction
(\ref{eq:stucorrelation}) which is independent of of microscopic details
of the measurement process.

There is yet another logical possibility one can entertain to avoid
the discrepancy pointed out above: That is that the particle positions
observed are not the $q_i$ but rather something else, some property
only of the wave function and the measurement device and independent
of the $q_i$. But seriously entertaining this possibility takes out
these ``true particle positions'' of any empirically founded scientific
theory. 

In conclusion it appears that the current situation is that in
set-ups such as those discussed in this paper involving more than
one instant of time either the predictions of the Bohmian particle
positions are incompatible with standard expectations or there are no
reliable predictions pending additional microscopic information or
alternatively detailed knowledge of all degrees of freedom in the
entire universe. The statement ``the Bohmian theory trivially makes
the same predictions as standard quantum mechanics because the
positions of all particles are distributed according to the quantum
mechanical probability distribution at all times'' seems to not be a
complete description in any case. This statement is only true if all
measurements are supposed to happen at the same time. But even if one
is happy with the ``all measurements are position measurements after
all, possibly of a pointer'' there is no justification for only
restricting to instantaneous measurements. And without this assumption
not only there is no basis for the claim that the Bohmian positions
agree with the $|\psi|^2$ distribution and thus with any other
interpretation. In fact, in this note we show this is manifestly wrong.

On the other hand, a restriction only to position variables at a
single instance of time would be a severe restriction that reduces
quantum physics to a subsector (a commutative subalgebra) which lacks
quantum properties as its states can be identified with classical
probability distributions (thanks to the
Gelfand-representation\cite{Gelfand}).

In any case, the system presented here makes quantitative predictions
in the conventional description which could easily be tested in an
actually possible (not just speculative and potentially impossible)
experiment which at best is a challenge for the Bohmian theory to
reproduce and for which a straight forward application of the guiding
equation and the identification of the $q_i$ with observable particle
positions yields a quantitatively different prediction as shown in Fig.~\ref{fig:twotimes}.

\acknowledgments I would like to thank J\"urg Fr\"ohlich and unnamed members of the
``Workgroup Mathematical Foundations of Physics'', Tim
Maudlin and Ward Struyve for helpful
discussions that helped to sharpen the points made in this paper. 

\section{Appendix: Frauchinger-Renner states}
A similar argument (although slightly more complex in the technical
details) can be made for a state like that investigated by Frauchinger
and Renner\cite{FR} (for an analysis from the point of view of the
Bohmian theory see \cite{LH}).

Here, we are dealing with two two-state systems. For the two-state Hilbert space we
use a basis $\downarrow/\uparrow$ or alternatively $\leftarrow =
(\downarrow-\uparrow)/\sqrt 2$ and $\rightarrow = (\downarrow +
\uparrow)/\sqrt 2$ as well as $\circlearrowleft=
(\uparrow+i\downarrow)/\sqrt 2$ and $\circlearrowright= (\uparrow -
i\downarrow)/\sqrt 2$.

On the bipartite system, we prepare the state
\begin{eqnarray*}
    \Psi_{FR} &=& \frac{\circlearrowright\otimes\circlearrowleft+(1+i) \circlearrowleft\otimes
  \leftarrow}{\sqrt 3}\\
&=&
\frac{\circlearrowleft\otimes\circlearrowright+\circlearrowright\otimes\circlearrowleft+i\circlearrowleft\otimes\circlearrowleft}{\sqrt
    3}\\
                            &=&
\frac{(1-i)\leftarrow\otimes\circlearrowright+(1+i)\rightarrow\otimes\circlearrowright+2(1+i)\leftarrow\otimes\circlearrowleft}{\sqrt
  {12}}\\
&=&
\frac{3\leftarrow\otimes
  \leftarrow+i\rightarrow\otimes\leftarrow+i\leftarrow\otimes
    \rightarrow+\rightarrow\otimes\rightarrow}{ \sqrt{12}}\\
  \end{eqnarray*}
From the third line, one can conclude that if one measures the first
particle in the state $\rightarrow$ the second particle will always in the
state $\circlearrowright$. From the second line, one can conclude that if one however
measures the second particle in the state $\circlearrowright$ the first particle
has to be in the state $\circlearrowleft$. Finally, if one however measures
the first particle in the state $\circlearrowleft$ the second particle will
always be in the state $\leftarrow$ as can be seen from the first line.

It would be tempting to combine these three observations to come to
the wrong conclusion that the combination
$\rightarrow\otimes\rightarrow$ will never be realised but the fourth
line shows that in fact it will be found in $1/12$ of the cases when
measuring both systems in the horizontal basis. The flaw is once more
to argue what would have happend if measured in the circular basis
(and assuming corresponding values) if
what one measures in fact is the horizontal basis. 

Now we have to translate this to the ``positions at different times''
language. We take two particles confined to their respective intervals
as before and focus on the subspace spanned by the ground state $g(x)$
and the
first excited state $e(x)$.

We represent $\uparrow$ and $\downarrow$ by
\begin{equation*}
  f(x) = \frac{g(x) + ie(x)}{\sqrt 2}\qquad b(x)= \frac{g(x) -
    ie(x)}{\sqrt 2}
\end{equation*}
and, to better match the ``if\dots the''-structure of the three
observations use as observables the projectors to the two halves of
the interval
\begin{equation*}
  P^\pm_t = \frac{1\pm\sigma_t(x)}2.
\end{equation*}
The wave function is then
\begin{equation*}
  \Psi_{FR}(x,y) = \frac{f(x) b(y) + b(x) f(y) + if(x) f(y)}{\sqrt 3}
\end{equation*}
for which
\begin{equation*}
  \langle \Psi_{FR}|P^-_{\pi/3}\otimes P^+_0|\Psi_{FR}\rangle = \frac 14 -
\frac 2{9\pi} -\frac {32}{27\pi^2}= 0.0591\ldots.
\end{equation*}
This can be read as the statement that if the first particle is in the
left half of the interval at time $\pi/3$, the second particle has
very likely been in the left half of its interval (since the
probability of finding it in the right half is very small). It would in
fact be $0$ if the factor of $8/3\pi$ would be absent in
(\ref{heisenberg}) which arises since multiplying the position
representation of $\sigma_0$ with the two lowest states not only
overlaps with the states we are interested in but also with higher
excited states and thus is not unitary in our two-dimensional Hilbert
sub-space of interest.

Furthermore, also
\begin{equation*}
  \langle \Psi_{FR}|P^+_{\pi/3}\otimes P^+_{\pi/3}|\Psi_{FR}\rangle =
  \frac 14 -\frac 4{9\pi} -\frac{16}{27\pi^2} = 0.0484\ldots
\end{equation*}
Therefore, at time $\pi/3$, if the second particle is in the right half of its
interval, it is highly likely that the first particle is in its left
half (as the probability of both of them being on the right is very
small).

Finally,
\begin{equation*}
\langle \Psi_{FR}|P^+_{0}\otimes P^-_{\pi/3}|\Psi_{FR}\rangle=\frac 14 -
\frac 2{9\pi} -\frac {32}{27\pi^2}= 0.0591\ldots.
\end{equation*}
which can be read as finding the first particle initially on the right
half of its interval strongly implies that at time $\pi/3$ the second
particle is on the right half.

Again, it is, however, wrong to combine these three conclusions to
find that virtually no pairs of particles can both be on the right
hand side as
with probability $1/12$, both particles are initially on the right
halves of their respective intervals.

The wrong conclusion is, however, reached if one assumes (as it is
done in the Bohmian interpretation) that at all times the particles
have actual positions and not just when they are measured.

Note that all these correlations would be confirmed by experiments
since none of the required measurements would require measuring the
position of a particle at more than one time. Once more, the only way
out of the paradoxical conclusion seems to be to assume that measuring
the position of one particle at some time ruins the predictability
of positions of the other particle from that time on. If that were the
case, any observation would invalidate quantum
mechanical predictions of other degrees of freedom in the universe for
later times because of the non-local nature of the Bohmian flow
equation where thanks to unknown entanglement the intervention of
measuring one degree of freedom would affect the time evolution of
other entangled degrees of freedom ruining predictability (where in
the standard quantum mechanical theory future predictions are not
affected as long as the two measuring operators do commute.

It should be noted that in fact all small probabilities above would
exactly be 0 if the matrix $\sigma$ were exactly the Pauli matrix and
would not have the fidelity factor $8/3\pi$ in front.

\bibliographystyle{JHEP}
\bibliography{GHZbohm}

\end{document}